\documentclass[prl,twocolumn, showpacs,preprintnumbers,amsmath,amssymb,superscriptaddress]{revtex4}

\usepackage{graphicx}

\begin{document}


\title {\bf 
Unconventional superconductivity in Na$_{0.35}$CoO$_{2}\cdot$1.3D$_{2}$O
and proximity to a magnetically ordered phase\/}
\author{Y.J.~Uemura}
\author{P.L.~Russo}
\author{A.T.~Savici}
\affiliation{Physics Department, Columbia University, 538 West, 120th Street,
New York, NY 10027, USA}
\author{C.R.~Wiebe}
\affiliation{Physics Department, Columbia University, 538 West, 120th Street,
New York, NY 10027, USA}
\affiliation{Department of Physics and Astronomy, 
McMaster University, Hamilton, Ontario L8S 4M1, Canada}%
\author{G.J.~MacDougall}
\author{G.M.~Luke}
\affiliation{Department of Physics and Astronomy, 
McMaster University, Hamilton, Ontario L8S 4M1, Canada}%
\author{M.~Mochizuki}
\author{Y.~Yanase}
\author{M.~Ogata}
\affiliation{Physics Department, University of Tokyo, Hongo, Tokyo 113-0033, 
Japan}
\author{M.L.~Foo}
\author{R.J.~Cava}
\affiliation{Chemistry Department and Princeton Materials Institute,
Princeton University, Princeton, NJ 08544, USA}%
\date{\today}
\begin{abstract}  
{ 
Muon spin relaxation ($\mu$SR) 
measurements on the  
new layered cobalt oxide superconductor Na$_{0.35}$CoO$_{2}\cdot$1.3H$_{2}$O
and its parent, non-superconducting compounds, have 
revealed unconventional nature of superconductivity 
through: (1) a small superfluid energy which implies
a surprisingly high effective mass of the charge carriers, approximately
100 times the bare electron mass; 
(2) the superconducting
transition temperature $T_{c}$ scaling with the 
superfluid energy following the correlations found in  
high-$T_{c}$ cuprate and some other two-dimensional
superconductors; (3) an anisotropic
pairing without broken time-reversal symmetry;
and (4) the proximity of a magnetically ordered insulating phase
at Na$_{0.5}$CoO$_{2}$ below $T_{N}$ = 53 K.}
\end{abstract}
%

\maketitle

\parskip 0cm

\narrowtext

Muon spin relaxation ($\mu$SR) measurements have been very effective
in demonstrating unconventional superconductivity in high-$T_{c}$
cuprate (HTSC) and organic superconductors.  The absolute
value of the measured penetration-depth $\lambda$
established correlations 
between $n_{s}/m^{*}$ (superconducting carrier density / effective mass)
and $T_{c}$ [1-3] which, together with the pseudogap behavior, suggest a 
formation of paired charge carriers 
occuring possibly at a temperature significantly 
higher than the condensation temperature $T_{c}$ [3,4].  
The temperature dependence of $\lambda$ indicated d-wave pairing 
symmetry and line nodes in the energy gap
[5,6].  Zero-field $\mu$SR studies 
revealed and elucidated static magnetic order 
in parent/relevant compounds of HTSC [3]. 

To these superconductors based on 
strongly correlated electrons, the recent discovery of superconductivity 
in Na$_{0.35}$CoO$_{2}$ intercalated with 1.3 H$_{2}$O [7]
has added a unique compound which has highly 2-dimensional (2-d) conducting 
planes of cobalt oxide in a triangular lattice structure
with geometrical spin frustration. The original idea of resonating
valence bonds was developed for this geometry [8], 
but no superconducting system in this geometry has been known before the 
new cobalt oxide compound.
Although extensive studies have been started [9-15],
detailed characteristics of this system are yet to be
demonstrated by conclusive experimental data sets.

We performed muon spin relaxation ($\mu$SR) measurements at 
TRIUMF in superconducting Na$_{0.35}$CoO$_{2}$ intercalated with 
1.3D$_{2}$O 
per formula unit, as well as in 
anhydrous Na$_{x}$CoO$_{2}$ with the Na concentration $x$ = 0.35, 0.5, 
and 0.64.  The samples were prepared at Princeton as described in earlier
reports [14,15], and pressed into disc-shaped
pellets with diameters of 6 mm.  Electron microscopy
of the pressed pellet samples indicates that the 2-d cobalt
oxide planes of the hydrated samples are essentially aligned.
The susceptibility ($\chi$) measurements showed 
superconducting $T_{c}$ = 4.2 K 
for the sample with D$_{2}$O. 
The superconducting sample and 
Na$_{0.35}$CoO$_{2}$, sensitive to air exposure, were transported 
to TRIUMF in sealed containers.  
Measurements 
were performed at T$\ge$25 mK using  
a dilution cryostat.         

We first describe Zero-field (ZF) $\mu$SR [16] 
studies of mangetic order in non-superconducting anhydrous
Na$_{x}$CoO$_{2}$.
Recent resistivity and susceptibility studies  
by Foo et al. [15] showed that the $x$ = 0.64 system can be
chracterized as the ``Curie-Weiss'' metal, $x$ = 0.35 as 
a ``paramagnetic'' metal, while $x$ = 0.5 exhibits 
a transition from a high-temperature
metal to low-temperature insulator at T = 53 K.
In Fig. 1(a), we show the ZF-$\mu$SR time spectra of these systems.   
In the $x$ = 0.5 system, the spectra above
T = 53 K show slow relaxation without oscillation, i.e., a line shape 
expected for systems with nuclear dipolar fields without
static magnetic order of Co moments.  Below T = 53 K, a clear
oscillation sets in, together with a rather fast damping.
Below T = 20-25 K, we see two frequencies beating.
Figure 1(b) shows the temperature 
dependence of these frequencies. 
The amplitude of the damping signal indicate that all the 
muons feel a strong static magnetic field below T = 53 K.
The 
static magnetic order sets in at the onset of a metal-insulator
transition, and the establishment of the second frequency
takes place at T = 20 K, which roughly corresponds to the ``kink''
temperature in the resistivity shown in the 
inset.  Although a conclusive picture requires
neutron scattering studies, it seems that one of 
two interpenetrating Co spin networks acquires a long-range order
below T = 53 K, followed by the other network establishing
long-range order below 20 K.  The spatial spin correlation 
should be antiferromagnetic (AF), since susceptibility shows
no divergence at T = 53 K [evidence against ferromagnetism], and 
the damping of the T = 25 mK data is significantly slower than
that of the Bessel function expected for the incommensurate spin-density-wave
(ISDW) states [17] [evidence against ISDW].  

We also confirmed the absence of static magnetic order in 
anhydrous Na$_{x}$CoO$_{2}$ with $x$ = 0.35 and 0.64, down to 
T = 25-35 mK, as shown in Fig. 1(a).  Static antiferromagnetic 
order was reported for $x$ = 0.75 - 0.9
by earlier $\mu$SR studies [18].  Together, the present data establish
a rather complicated evolution of the magnetic ground states from 
paramagetic (PM) ($x$ = 0.35) to AF (0.5) to PM (0.64) to AF (0.75) to ISDW 
(0.9),
with increasing $x$.
The $\sim$ 2 MHz frequency in the $x$ = 0.5 system is close to
$\sim$ 3 MHz in $x$ = 0.75,
suggesting that the ordered moment sizes in these systems 
are of comparable magnitudes. The existence of an insulating magnetic
state in the vicinity of superconductivity resembles the case 
in the cuprates.  

Intercalation of H$_{2}$O or D$_{2}$O into the Na$_{x}$CoO$_{2}$
yields superconducting systems in a rather narrow range of $x$ [14].
ZF-$\mu$SR is a powerful tool to detect a static magnetic field due to
the particular superconducting pairing states associated with 
Time-Reversal-Symmetry-Breaking (TRSB), as shown in the case of 
Sr$_{2}$RuO$_{4}$ [19].  We observed Gaussian damping of the muon asymmetry
in ZF-$\mu$SR of 
Na$_{0.35}$CoO$_{2}\cdot$1.3D$_{2}$O.  This damping is due to 
nuclear dipolar fields, and the Gaussian shape comes from the 
initial decay of the Kubo-Toyabe function for nuclear dipolar 
broadening [16].  Since the recovery of this function was missing
in our observable time range (up to 8 $\mu$s), we fitted this
damping with the simple Gaussian function $\exp(-\sigma^{2}t^{2}/2)$.  
As shown in Fig. 2(a), the relaxation rate 
$\sigma$ in ZF is independent of temperature between T = 6 K and 
T = 25 mK.  The arrows with ``TRSB'' indicate the expected changes of 
$\sigma$ for the TRSB fields having random directions and Gaussian
distribution of width (RMS second moment) 1 G and 2 G,
respectively, added quadratically to the nuclear dipolar fields. 
Our results rule out the existence of a TRSB field above the 1 G level.
This is consistent with an earlier report [10], yet we extended
the temperature range from 2 K to 25 mK.
On a triangular lattice, d-wave
pairing has coexisiting real and imaginary parts, resulting in a 
TRSB field.  The present data sets a rather severe constraint to 
the d-wave pairing cases.     
 
$\mu$SR data in transverse external fields (TF) reflect field
broadening due to the flux vortex lattice in type-II superconductors,
from which one can derive the magnetic field penetration depth 
$\lambda$ [3,5].  We performed TF-$\mu$SR measurements
in superconducting samples intercalated with D$_{2}$O [Fig. 2 (a)(b)],
with the external field TF = 200 G applied
perpendicular to the aligned CoO planes. 
Figure 2(b) shows the muon 
spin relaxation rate, fitted to the Gaussian damping 
$\exp(-\sigma^{2}t^{2}/2)$, 
with $\sigma_{n}$ indicating the average relaxation rate in 
the normal state.  If the observed change in TF=200 G were due to 
a mechanism sensitive to ZF-$\mu$SR, we would have observed a change
of the ZF relaxation rate to the level indicated by the ``TF'' arrow
in Fig. 2(a).  Thus, we proceed our discussion by assuming that the
increase of $\sigma$ in TF in Fig. 2(b) is solely due to  
the in-plane penetration depth $\lambda_{ab}$.  
By quadratically subtracting $\sigma_{n}$
from the observed relaxation rate $\sigma_{exp}$, we obtained
the relaxation rate $\sigma_{sc}$ due to superconductivity
as shown in Fig. 3.  In separate measurements (not shown), we 
found essentially no dependence of $\sigma_{sc}$ on 
TF in the range between 100 G and 2 kG, which assures no 
involvement of 2-d pancake vortex formation [5].      

In Fig. 3, we compared the temperature dependence of 
$\sigma_{sc}(T) \propto \lambda^{-2}$ of Na$_{0.35}$CoO$_{2}\cdot$1.3D$_{2}$O
with various models, in a fit of 16 data points 
with $\sigma(T=0)$ as a free parameter.
The observed results clearly disagree
with curves of the two-fluid model (normalized chi square NCS=3.51) and 
s-wave BCS weak-coupling model (NCS=1.75, Durbin-Watson value of
a normalized residual error correlations DW=1.11).
Comparison with the scaled $\mu$SR results from YBCO [5] yields
NCS=1.39 and DW=1.59, showing a rather poor agreement yet in 
a statistically acceptable range.  For a 5\%\ confidence level,
a model with NCS$>$1.666 or DW$<$1.1 or DW$>$2.9 should be rejected,
1.1$<$DW$<$1.37 or 2.63$<$DW$<$2.9 is inconclusive, while 1.37$<$DW$<$2.63
is comfortably acceptable.

For the cobalt oxide superconductors, several authors proposed
f-wave models [13,20], which have a particular matching with 
the symmetry of triangular lattice. 
In Fig. 3, we also show a theoretical curve for an 
f-wave pairing, obtained by
using a tight-binding fit of the LDA band calculation [13] 
and by assuming a separable effective interaction supporting
a simple f-wave order parameter. 
In the present system, there is a large 
Fermi surface around the $\Gamma$ point as
well as six small hole-pockets near the K points.
The line in Fig. 3 represents 
a case where nodes of f-wave symmetry exist only
on the large Fermi surface and not on the six hole-pockets,
while the order parameter on each Fermi 
surface has the same maximum value.  
This f-wave model gives a good agreement with 
the observed data with NCS=1.19 and DW=2.34.  

These results rule out a fully isotropic energy gap.
Before concluding a particular pairing symmetry, however,
one has to test various other models with/without the possible effect of
impurities.  In ref. [11] the authors discussed 
anisotropy of the energy gap 
based on TF-$\mu$SR data with a few temperature points below $T_{c}$.
Our finding of an anisotripic energy gap is 
consistent with earlier reports of a power-law T-dependence of 
the NMR relaxation rate $1/T_{1}$ as well as with the T-independent
Knight shift of NMR [9] and $\mu$SR [10] below $T_{c}$. 
      
The penetration depth $\lambda$ is related to the 
superconducting carrier density $n_{s}$ divided by the 
effective mass $m^{*}$ as
$\sigma(T) \propto \lambda^{-2} \propto
[4\pi n_{s}e^{2}/m^{*}c^{2}][1/(1+\xi/{\it l\/})]$,
where $\xi$ is the coherence length and {\it l\/} denotes
the mean free path.  At this moment, it is difficult to 
prove the clean limit situation $\xi <<$ {\it l\/} for
the cobalt oxide superconductor, 
due to the lack of high-quality superconducting single crystals
necessary to estimate the in-plane values of $\xi$ and 
{\it l\/}.  In the following, we proceed the discussion of the 
superfluid energy scale $n_{s}/m^{*}$ by assuming the clean-limit,
in view of an excellent conductivity in anhydrous Na$_{0.31}$CoO$_{2}$
crystals [15] and high $H_{c2}$ values in polycrystalline superconducting
specimens [21].  

Derivation of the absolute values of $\lambda$ and $n_{s}/m^{*}$
is subject to modeling of flux vortex lattice line shapes, observed
functional forms of field distribution, and angular averaging in the 
polycrystal samples.  Based on the results of $\mu$SR measurements
on c-axis aligned YBCO [22] and numerical works [23], we have adopted
the conversion factor for polycrystal to aligned samples 
$\sigma_{aligned} \sim 1.4 \sigma_{poly}$ to account for the effect of 
applying the TF perpendicular to the conducting planes of highly
2-d superconductors. For $\sigma$ to $\lambda$
conversion $\lambda = A /\sqrt{\sigma}$, we have adopted a factor
$A$ = 2,700 [\AA ($\mu$s)$^{1/2}$] for the Gaussian width $\sigma$.  
With these conversion factors, the values of 
$\lambda_{ab}$ of polycrystalline samples of underdoped YBCO  
with $T_{c} \sim 60$ K [1] agree well with the value 
obtained using a single crystal specimen with comparable $T_{c}$
in a more accurate line-shape analysis [5].  
The above factor $A$ gives $\lambda$ = 7,200 \AA\ 
for the in-plane penetration depth of the cobalt oxide system at 
$T\rightarrow 0$.


For highly 2-d superconductors, it is also 
interesting to study correlations between $T_{c}$ and the 
2-d superfluid density $n_{s2d}/m^{*}$ which 
can be obtained by multplying $\sigma_{aligned}$ with the average
distance $c_{int}$ of conducting planes. 
Figure 4 shows such a comparison, including the cuprates [3,22], 
alkali-doped (Hf/Zr)NCl with/without intercalation of THF
(tetrhydrofuran) [24], and organic 2-d superconductors based on 
(BEDT-TTF) salts [6].  All the data points are taken using 
single crystal or aligned samples with TF perpendicular
to the conducting planes, while
``cuprate'' lines represent polycrystal results [1,3] after
the factor 1.4 correction.  We find that $T_{c}$ of all these
2-d superconductors could have a common relationship to the 2-d superfluid
density $n_{s2d}/m^{*}$, which can be converted into corresponding
2-dimensional Fermi energy as given in the lower horizontal axis.

Based on correlations between $T_{c}$ and the superfluid density
in the cuprates, Emery and Kivelson [4] proposed a picture in which 
$T_{c}$ is determined by phase fluctuations in the argument
essentially identical to the Kosterlitz-Thouless (KT) theory [25].
In KT transitions, $T_{c}$ and the superfluid density at the 
transition temperature $T_{KT}$ are related with a universal 
system-independent relationship, which is shown by the $T_{KT}$ 
line in Fig. 4.  In Fig. 4, most of the points lie at about a factor 2-3 away
in the horizontal axis from the $T_{KT}$ line, which implies
that the superfluid density undergoes about a 2-3 times reduction from 
the $T=0$ value to the value near $T_{c}$ where phase fluctuations may
destroy 3-d superconductivity.  In the cuprates, this reduction
could be related to excitations of nodal quasi particles, or 
classical thermal fluctuations, or some elementary excitations.
Further studies for the origin of the T-dependence of $\sigma(T)$ 
could provide a key to understanding the correlations shown in Fig. 4.

If we assume the charge carrier density to be equal to the Na concentration,
we obtain the in-plane effective mass of the cobalt-oxide superconductor
to be about 100 times the bare electron mass $m_{e}$.  
A similar estimate for $m^{*}$
was given in ref. [11].
The heavy mass can be expected for strongly correlated carriers in a triangular
lattice [12].    The high effective mass 
is consistent with the electronic
specific heat $C/T \sim 12$ [mJ/mole K$^{2}$] of the superconducting
cobalt oxide [26] just above $T_{c}$.  This value can be compared to 
$\sim 2$ [mJ/mole K$^{2}$] of YBa$_{2}$Cu$_{3}$O$_{7}$ [27].
After normalizing the values to a unit sheet area of conducting 
planes, $C/T$ for the cobalt oxide becomes about 25 times larger
than that for YBCO.  In the non-interacting 2-d Fermi gas, $C/T$ is
proportional to $m^{*}$ but independent of carrier density.  Thus, 
within this approximation, we expect $m^{*}$ of cobalt oxide 
to be 25 times that of the cuprates.

In conclusion, we have shown that the cobalt oxide superconductors
have an anisotropic energy gap
and a heavy effective mass $m^{*} \sim$ 100$m_{e}$,
without a TRSB field (limit given as 1 G).
We established the
existence of an antiferromagnetic insulating compound in the vicinity
of the superconducting cobalt-oxide system without magnetic order,
which suggests the possible involvement of magnetism in the superconducting
mechanism.  

The work at Columbia has been supported by the NSF 
DMR-0102752 and CHE-0117752 (Nanoscale Science and Engineering
Initiative), at Princeton by NSF DMR-0213706 (MRSEC)
and by the DOE DE-FG02-98-ER45706, and   
at McMaster by
NSERC and the CIAR (Quantum Materials Program).
We acknowledge F.D. Callaghan and J.E. Sonier for technical assistance. 

%
%
%
%

\vskip 1.0 truecm

\begin{figure}[h]

\begin{center}
\includegraphics[angle=270,width=3.30in]{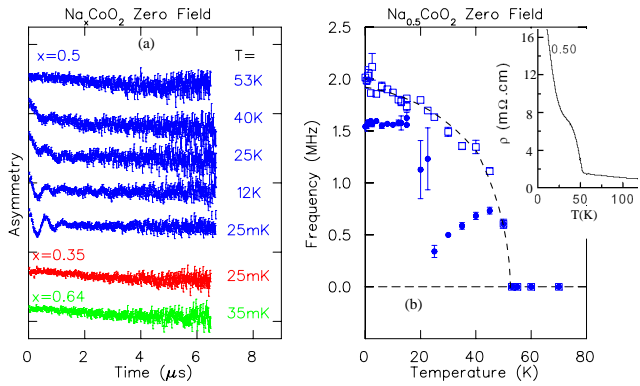}


\caption{\label{Figure 1.}
(a) Muon spin relaxation time spectra observed in zero field 
in anhydrous Na$_{x}$CoO$_{2}$ with $x$ = 0.50, 0.35 and 0.64.
(b) The muon spin precession frequency observed in the $x$ = 0.5 system,
shown with the resistivity in the inset [15].  
}

\end{center}
\end{figure}


\begin{figure}[h]

\begin{center}

\includegraphics[width=2.25in]{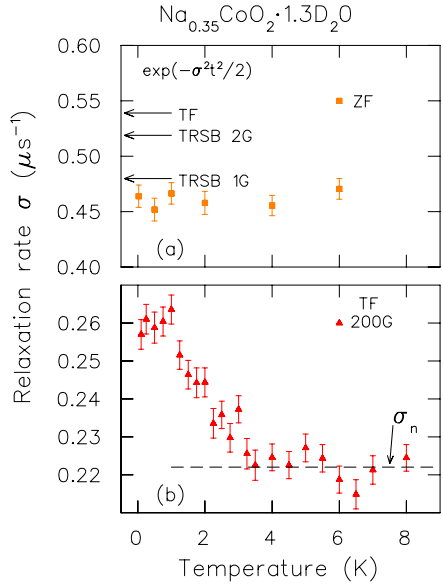}


\caption{\label{Figure 2.} 
Muon spin relaxation rate observed in 
a superconducting 
specimen of 
Na$_{0.35}$CoO$_{2}\cdot$1.3D$_{2}$O, which   
has $T_{c}(\chi)$ = 4.2 K from susceptibility $\chi$.
(a) shows results in zero field, while (b) in TF = 200 G
applied perpendicular to the conducting planes.  
See text for the arrows in (a) and
$\sigma_{n}$ in (b).} 

\end{center}
\end{figure}


\begin{figure}[h]

\begin{center}

\includegraphics[angle=90,width=2.7in]{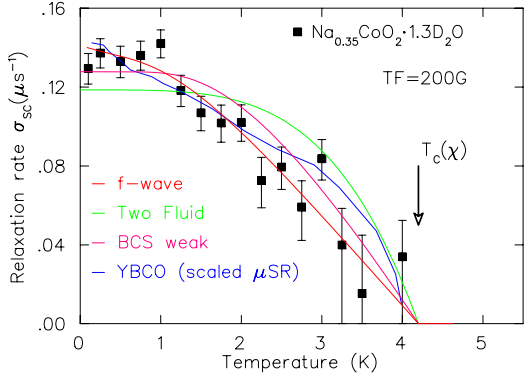}


\caption{\label{Figure 3.}Muon spin relaxation rate $\sigma_{sc}(T)$ due to superconductivity
in Na$_{0.35}$CoO$_{2}\cdot$1.3D$_{2}$O,
with TF = 200 G applied perpendicular to the aligned conducting planes,
obtained by quadratic subtranction of $\sigma_{sc}^{2} = \sigma_{exp}^{2}
-\sigma_{n}^{2}$, where $\sigma_{exp}$ and $\sigma_{n}$ are shown in
Fig. 2(b).  The results are compared with fits to 
several models
and the scaled plot of $\mu$SR results on 
YBa$_{2}$Cu$_{3}$O$_{6.95}$ (YBCO) [5].} 

\end{center}
\end{figure}

\begin{figure}[h]

\begin{center}

\includegraphics[width=2.5in]{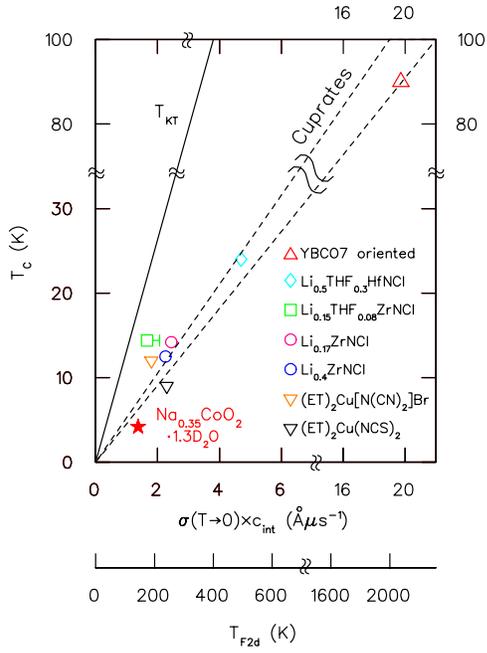}


\caption{\label{Figure 4.} 
A comparison of highly 2-d superconductors in a plot of 
$T_{c}$ versus 
$\sigma_{sc}(T=0)$,
multiplied by the average interlayer
distance $c_{int}$ of the conducting planes. 
Data points are from
aligned pellet or single-crystal specimens
[6,22,24], 
while the dotted lines are from 
ceramic specimens of YBCO [1-3]
after a factor 1.4 correction.  
For $\sigma \times c_{int} \propto n_{s2d}/m^{*}$,
we show the
corresponding Fermi temperature
$T_{F2d}$ of the 2-d electron gas.}

\end{center}
\end{figure}
\end{document}